\documentclass[twocolumn, switch]{article}

\usepackage{preprint}

\usepackage{amsmath, amsthm, amssymb, amsfonts, bm}

\usepackage[numbers,square]{natbib}
\usepackage[utf8]{inputenc}	
\usepackage[T1]{fontenc}	
\usepackage{xcolor}		
\usepackage[colorlinks = true,
            linkcolor = purple,
            urlcolor  = blue,
            citecolor = cyan,
            anchorcolor = black]{hyperref}	
\usepackage{booktabs} 		
\usepackage{nicefrac}		
\usepackage{microtype}		
\usepackage{lineno}		
\usepackage{float}		
\usepackage{graphicx}  
\usepackage{subfigure}
\usepackage{wrapfig}
\usepackage{multirow}

\usepackage{fancyhdr}
\usepackage{longtable}
\usepackage{parskip}
\usepackage{dcolumn}   

\usepackage{soul}      

\newcommand{\pic}[4]{\begin{center} \includegraphics[scale=#1]{#2}\label{#3}\caption{#4}\end{center}}

\newcommand{\pwisein}{\left\{ \begin{array}{ll}}
\newcommand{\pwiseout}{\end{array}\right.}

\renewcommand{\det}[1]{\mathrm{det}\left( #1 \right)}

\usepackage{newfloat}
\DeclareFloatingEnvironment[name={Supplementary Figure}]{suppfigure}
\usepackage{sidecap}
\sidecaptionvpos{figure}{c}

\usepackage{titlesec}
\titlespacing\section{0pt}{12pt plus 3pt minus 3pt}{1pt plus 1pt minus 1pt}
\titlespacing\subsection{0pt}{10pt plus 3pt minus 3pt}{1pt plus 1pt minus 1pt}
\titlespacing\subsubsection{0pt}{8pt plus 3pt minus 3pt}{1pt plus 1pt minus 1pt}

\usepackage{xr}
\usepackage{pdfpages}

\title{The effect of the A-site cation on the phase transition temperature of metal halide perovskites}

\usepackage{authblk}

\author[1,2]{Tom Braeckevelt}
\author[1]{Sander Vandenhaute}
\author[1]{Sven M.\ J.\ Rogge}
\author[2]{Johan Hofkens}
\author[1]{Veronique Van Speybroeck$^{\dagger,}$}

\affil[1]{Center for Molecular Modeling, Ghent University, Technologiepark 46, Zwijnaarde 9052, Belgium}
\affil[2]{Department of Chemistry, KU Leuven, Celestijnenlaan 200F, Leuven 3001, Belgium}

\begin{document}

\twocolumn[ 
  \begin{@twocolumnfalse} 

\maketitle

\begin{center}
  $\dagger$ Email: \texttt{Veronique.VanSpeybroeck@UGent.be}
\end{center}

\vspace{0.35cm}

\begin{abstract}
A key challenge for the practical application of metal halide perovskites (MHPs) is the instability of the desired perovskite phase relative to the optically non-active $\delta$ phase.
To determine the phase stability, we previously developed a procedure to compute the harmonic free energy as a function of temperature, which was suited for CsPbI$_3$ but fails when Cs is replaced by organic cations due to their rotational freedom.
Herein we propose a multistep thermodynamic integration (TI) approach that corrects the harmonic free energy to obtain the Gibbs free energy. 
Given the abundance of local minima in these materials, we employ replica exchange to prevent simulations from getting trapped, while introducing an intermediate potential energy surface to improve convergence and reduce computational cost. 
Benchmarking energy and forces from different exchange-correlation functionals and dispersion methods against high-level RPA+HF calculations identifies PBE+D3(BJ) as the best trade-off between accuracy, computational efficiency, and precision. 
To perform molecular dynamics simulations within the TI framework, it was necessary to train a machine learning potential using the MACE architecture on \textit{ab initio} data calculated with density functional theory. 
Our results show that, for all three materials, the free energy difference between the $\gamma$ and $\delta$ phases exhibits a very similar temperature dependence. 
This suggests that phase stability is primarily governed by differences in ground-state energy, rather than by material-specific thermal effects.
Beyond these three materials, our methodology provides a robust framework for investigating the phase behavior of other MHPs, paving the way for the discovery of more stable perovskites.
\end{abstract}
\vspace{0.35cm}
 
  \end{@twocolumnfalse} 
]

\section{Introduction}
\begin{figure*}
	\centering
	\includegraphics[width=1\linewidth]{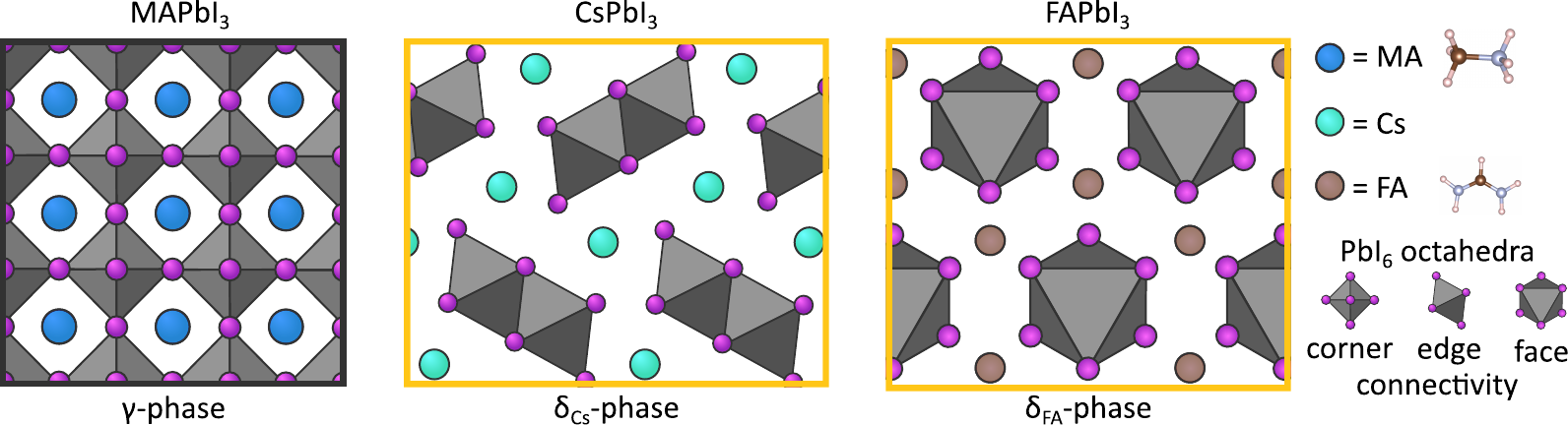}
	\caption{Schematic representation of the most stable phase for MAPbI$_3$, CsPbI$_3$, and FAPbI$_3$ at room temperature.}
	\label{FigPhases}
\end{figure*}

Over the past decade, metal halide perovskites (MHPs) with the general composition ABX$_3$ have garnered significant interest for a variety of optoelectronic applications.\cite{Kim2020,Zhang2021,Dong2023} 
However, many MHPs suffer from low stability and readily degrade under certain external conditions.\cite{Xiang2021}
In particular, widely studied MHPs such as CsPbI$_3$ and FAPbI$_3$ (FA = formamidinium) exhibit a perovskite phase (also referred to as the black phase or, depending on its symmetry, as the $\alpha$, $\beta$, or $\gamma$ phase) that is metastable at room temperature. 
Over time, this perovskite phase transitions into a non-perovskite phase, often called the yellow phase or $\delta$ phase, which significantly deteriorates the optoelectronic properties.
Interestingly, substituting the A-site cation from Cs$^+$ to FA$^+$ for APbI$_3$ MHPs leads to another $\delta$ phase that is stable at room temperature, while using methylammonium cations (MA$^+$) leads to a stable perovskite phase at room temperature.\cite{Stoumpos2013} 
Fig.\ \ref{FigPhases} illustrates the different PbI frameworks of the two $\delta$ phases and the $\gamma$ phase, introducing a nomenclature to distinguish the two $\delta$ phases: $\delta_{\text{Cs}}$ and $\delta_{\text{FA}}$.

In recent years, significant effort has been dedicated to stabilizing the perovskite phase across various MHPs. 
Experimental studies have shown that this can be achieved through compositional and strain engineering, surface functionalization, and crystal size control.\cite{Steele2019,Steele2021,Chen2019stab,Li2018,Lifaming2018,Wu2019,Dirin2016,Lu2018} 
However, the underlying driving forces behind these stabilizing effects remain poorly understood.
To address this gap, we computationally predict the stability of the perovskite phase, allowing us to gain molecular-level insights and identify the key stabilizating factors.\cite{Zhang_2018_stabMHPs, Arora2022} 
In previous work, we successfully predicted the transition temperature of CsPbI$_3$.\cite{Braeckevelt2022_CsPbI3} 
To further elucidate the A-site cations influence on phase stability at room temperature, we extend our approach to FAPbI$_3$ and MAPbI$_3$.
However, a key challenge arises: the organic cations possess rotational degrees of freedom, which significantly affect the free energy landscape. 
Our previous methodology relied on the harmonic approximation, which becomes inadequate when rotational motions are significant.\cite{Kapil2019_methods_anh_F}

To correct for the errors introduced by the harmonic approximation, we employ a combination of thermodynamic integration (TI) methods, following the approach outlined by Cheng and Ceriotti.\cite{Cheng2018_PRBv97}
TI has been successfully applied to various systems, including the thermodynamics of water,\cite{Cheng2019_waterTI} metal-organic frameworks,\cite{Rogge2015_MOFTI} nitrogen,\cite{Meijer1990_N_TI} proton hopping and adsorption in zeolites,\cite{Bucko2020_phop_TI, Amsler2021_adsZeo_TI} and multicomponent alloys.\cite{Grabowski2019_TIMLP}
TI states that the change in free energy along a given path is equal to the ensemble average of the derivative of the potential energy along that path.
The primary challenge lies in achieving convergence of these ensemble averages, which typically demands significant computational resources. 
As a result, TI has traditionally been applied to relatively simple model systems\cite{Frenkel1984_TIhardspheres, Broughton1983_TI_LJ} or studies relying on classical force fields.\cite{Li2017_TI_FF, Rogge2015_MOFTI}
Unfortunately, the accuracy of TI is highly dependent on the quality of the force field.\cite{Li2017_TI_FF} 

Recently, machine learning potentials (MLPs) have been employed in TI calculations to achieve \textit{ab initio} accuracy at a significantly lower computational cost.\cite{Cheng2019_waterTI, Grabowski2019_TIMLP, Fukushima2019_TI_MLP, Fransson2023_TI} 
This is necessary because the computational cost of \textit{ab initio} methods is prohibitive for the extensive sampling required in TI.
Message-passing equivariant MLPs such as NequIP\cite{Batzner2022_nequip} and MACE\cite{Batatia2022_MACE} combine high accuracy with a low data requirement.
This enables training on more accurate \textit{ab initio} data and broadens the range of materials and phases that can be studied.
To this end, we will evaluate the accuracy of various \textit{ab initio} methods, assessing both energy and force predictions by benchmarking them against RPA+HF.

An additional challenge arises when studying flexible materials such as MHPs.\cite{Bechtel2019} 
The phase space associated with a stable phase may contain energy barriers that are unlikely to be overcome during molecular dynamics (MD) simulations at lower temperatures.\cite{Leguy2015, Fransson2023_fesMHPs, Dutta2025, Carnevali2025} 
To avoid the system getting trapped into a local minimum, we apply replica exchange (REX) \cite{SUGITA1999_REX_NVT} assuming that the MD at the highest temperature can frequently cross those barriers. 
REX facilitates Monte Carlo (MC) swaps between MD runs at different temperatures, pressures, or Hamiltonians,\cite{Earl2005_REXreview} allowing structural information sampled at high temperatures to be transferred to lower temperatures. 
REX has been shown to efficiently explore phase space at lower temperatures, where conventional MD simulations would otherwise remain stuck in local minima.\cite{Earl2005_REXreview}
Due to its effectiveness, REX has been applied to various systems, including protein folding,\cite{SUGITA1999_REX_NVT} phase transitions in metal-organic frameworks,\cite{Demuynck2018_tica} and cation distributions in zeolites.\cite{Beauvais2004_REX}

This paper is structured as follows: First, we outline the theory and computational details of our revised methodology. 
Next, we evaluate the accuracy of different \textit{ab initio} methods. 
Finally, we apply our revised approach to CsPbI$_3$, FAPbI$_3$, and MAPbI$_3$, obtaining free energy differences between phases as a function of temperature, from which we determine the transition temperatures.

\section{Methodology}
\label{Methodology_section}
Thermodynamic integration (TI) is computationally expensive. 
To accelerate calculations, machine learning potentials (MLPs) are trained using MACE.\cite{Batatia2022_MACE} 
MACE is a message-passing neural network with equivariant features that significantly enhances both the accuracy and data efficiency of the model.\cite{Batzner2022_nequip} 
By extending two-body message passing to many-body interactions, MACE reduces the required number of interaction layers while maintaining similar accuracy, thereby lowering computational costs.\cite{Batatia2022_MACE}

We train MACE on \textit{ab initio} data generated using VASP\cite{VASP1,VASP2} with the PBE+D3(BJ) exchange-correlation (XC) functional.\cite{PBE,D3BJ}  
The PBE+D3(BJ) XC functional was chosen as it closely approximates RPA+HF energies and forces while requiring significantly fewer computational resources.\cite{Chen2017,Ramberger2017_RPAforces}
Further details on \textit{ab initio} data generation and MLP training are provided in Sec.\ \ref{Computational_details}.

TI accounts for anharmonicities in free energy calculations without assuming a specific form of the potential energy surface (PES).\cite{Frenkel_and_Smit} 
In general, TI computes the Helmholtz free energy, $F$, difference between two states within a given simulation cell, $\textbf{h}$, and temperature, $T$, by integrating the change in free energy along a path connecting states $a$ and $b$:
\begin{equation}
	\Delta_{a\rightarrow b}^\text{TI} F(\textbf{h},T) = \int_{a}^{b} \frac{\partial F(\lambda,\textbf{h},T)}{\partial \lambda} d\lambda = \int_{a}^{b} \Big\langle \frac{\partial U(\lambda)}{\partial \lambda} \Big\rangle_{\lambda,\textbf{h},T} d\lambda.
	\label{TI_general}
\end{equation}
For a system with $N$ atoms, the Helmholtz free energy is equal to $-k_\text{B} T \ln\left(\int{} e^{-\frac{U (\lambda, \vec{q})}{k_\text{B} T}} d\vec{q} \right)$, where $k_\text{B}$ is the Boltzmann constant and $\vec{q}$ a 6N-dimensional vector in phase space. 
If $\lambda$ is independent of temperature, then the free energy derivative simplifies to the ensemble average of the potential energy derivative, as shown in Eq.\ \ref{TI_general}. 
When $\lambda$ represents the temperature, the free energy difference between two temperatures can be computed as:
\begin{equation}
	\begin{aligned}
		&\Delta_{T_1\rightarrow T_2}^\text{TI}F(\textbf{h},T_2) = F(\textbf{h},T_2) - F(\textbf{h},T_1) = \\ 
		&\left(\frac{T_2 - T_1}{T_1}\right) F(\textbf{h},T_1) - T_2 \int_{T_1}^{T_2} \frac{\langle U\rangle_{\textbf{h},T'} + \frac{3N-3}{2} k_\text{B}T'}{T'^2} dT'.
	\end{aligned}
	\label{TemTI}
\end{equation} 
A similar expression holds for the Gibbs free energy, $G$, by replacing the potential energy, $U$, with the enthalpy, $H$.\cite{Cheng2018_PRBv97} 

The parameter $\lambda$ can represent a physical coordinate-dependent path between two states,\cite{Ciccotti2005,Amsler2023} or it can serve as a hypothetical interpolation variable that gradually transforms one potential energy surface (PES) into another.\cite{Jaykhedkar2024,Cheng2019_waterTI,Grabowski2019_TIMLP,Fukushima2019_TI_MLP}
For instance, starting from a harmonic PES, $U_\text{harm}$, we can linearly interpolate toward the MLP PES, $U_\text{MLP}$, by defining the path:
\begin{equation}
	U(\lambda)= U_\text{harm} + \lambda \left(U_\text{MLP} - U_\text{harm}\right) 
	\label{eq_intPES}
\end{equation}
for $0\leqslant\lambda\leqslant{} 1$. The Helmholtz free energy of the MLP PES can then be computed via thermodynamic integration (TI) as:
\begin{equation}
	\begin{aligned}
		\Delta_{\text{harm}\rightarrow \text{MLP}}^\text{TI}F_\text{MLP}(\textbf{h},T) & = F_\text{MLP}(\textbf{h},T) - F_\text{harm}(\textbf{h},T) \\
		&= \int_{0}^{1} \Big\langle U_\text{MLP} - U_\text{harm} \Big\rangle_{\lambda,\textbf{h},T_0} d\lambda.
	\end{aligned}
	\label{lambdaTI}
\end{equation} 
The Helmholtz free energy of the harmonic PES, $F_\text{harm}(\textbf{h},T)$, can be determined analytically:\cite{Peters2019} 
\begin{equation}
	F_\text{harm}(\textbf{h},T) = E_\text{GS}(\textbf{h}) + k_\text{B}T \sum_{i=1}^{3N-3}\ln \left(\frac{h \nu_i(\textbf{h})}{k_\text{B}T}\right),
	\label{HarmFreeEnergy}
\end{equation}
where $ E_\text{GS}(\mathbf{h})$ is the ground-state energy and $\nu_i(\textbf{h})$ are the frequencies. 
Here, $h$ is Planck’s constant and $N$ is the number of atoms in the system.

Finally, since the previous simulations are performed at fixed simulation cell $\mathbf{h}$, we convert the resulting Helmholtz free energy to the Gibbs free energy at pressure $P$ by applying the correction
\begin{equation}
	\begin{aligned}
	\Delta_{\mathbf{h}\rightarrow P} G(P,T) & = G(P,T) - F(\mathbf{h},T) \\ 
	& = P\det{\mathbf{h}} + k_\text{B}T \ln \rho(\mathbf{h}|P,T),
	\end{aligned}
	\label{HelmToGibbs}
\end{equation}
where $\rho(\mathbf{h}|P,T)$ is the probability of observing the simulation cell $\mathbf{h}$ in an MD simulation performed at constant pressure and temperature.\cite{Cheng2018_PRBv97} 
As discussed in Sec.\ \ref{workflow}, the probability distribution $\rho(\mathbf{h}|P,T)$ is approximated by $\rho(V|P,T)$, since converging the full six-dimensional distribution proved computationally unfeasible.

\begin{figure}[h]
	\centering
	\includegraphics[width=1.0\linewidth]{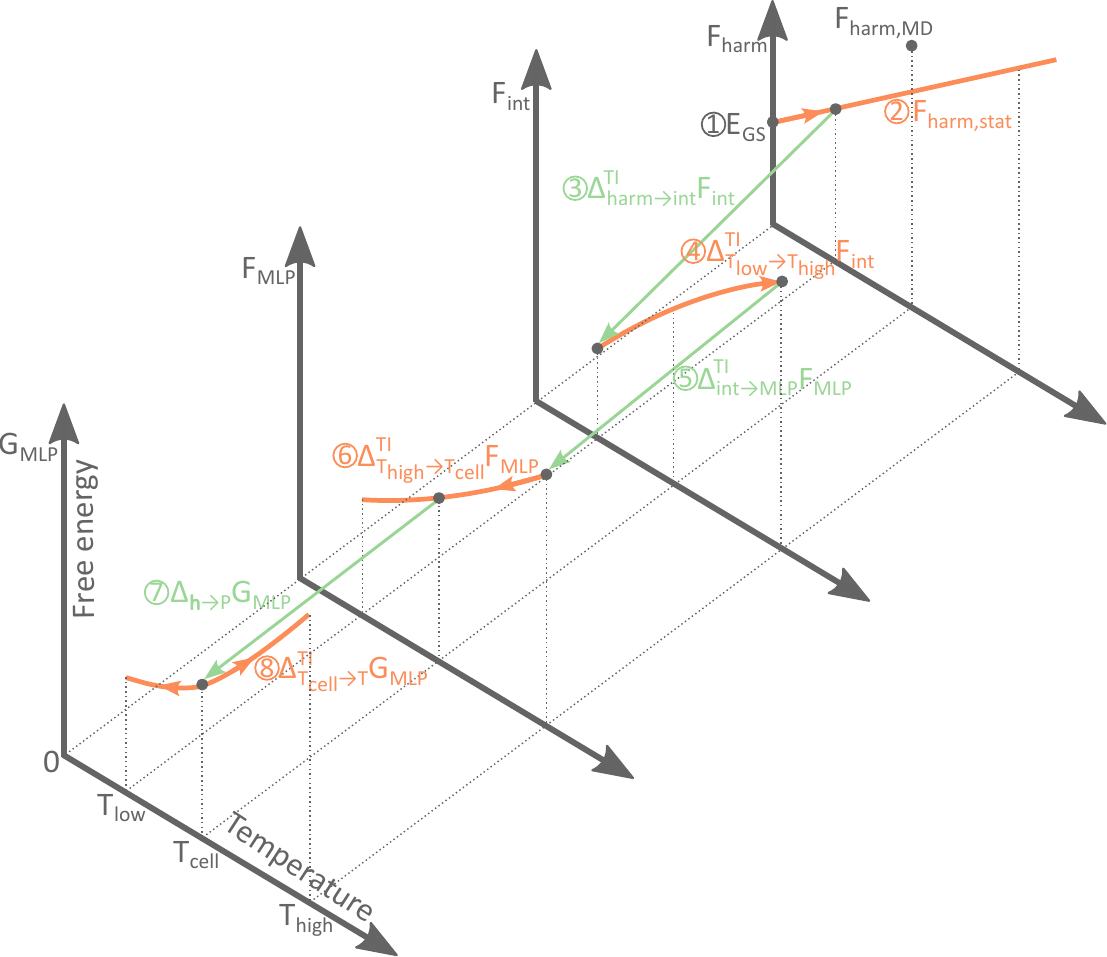}
	\caption{Visualization of the free-energy contributions required to obtain the Gibbs free energy as a function of temperature. 
		Starting from the ground-state energy, $E_\text{GS}$, we calculate the static harmonic free energy, $F_\text{harm,stat}$. 
		The Helmholtz free energy is evaluated at the average cell $\mathbf{h}$ obtained from an NPT MD simulation at temperature $T_\text{cell}$ and pressure $P$, i.e., the pressure at which the Gibbs free energy is desired. 
		A TI contribution at $T_\text{low}$ corrects for anharmonic effects, yielding $F_\text{int}(\mathbf{h},T_\text{low})$, followed by a TI contribution along a temperature path to $T_\text{high}$ and a TI correction from the intermediate to the MLP PES, giving $F_\text{MLP}(\mathbf{h}, T_\text{high})$. 
		To obtain $G_\text{MLP}$, a TI contribution is first performed from $T_\text{high}$ to $T_\text{cell}$, followed by the correction $\Delta_{\mathbf{h}\rightarrow P}$, and finally a TI contribution from $T_\text{cell}$ to any temperature $T$. 
		The previously reported $F_\text{H,MD}$ is shown for comparison.\cite{Braeckevelt2022_CsPbI3}}
	\label{Free_energy_drawing}
\end{figure}

As shown in Fig.\ \ref{Free_energy_drawing}, we only need these (TI) contributions to calculate the Gibbs free energy as a function of temperature. 
Moreover, MLPs allow TI corrections to converge with \textit{ab initio} accuracy.\cite{Jaykhedkar2024,Reinhardt2021} 
However, this straightforward approach led to convergence issues for MHPs. 
Therefore, more advanced sampling techniques than standard MD are required, and an intermediate PES, $U_\text{int}$, is introduced, as discussed in more detail below.

We first explain the core issue. 
Namely, these difficulties arise in systems where local minima are separated by high energy barriers, limiting phase-space exploration.
In such cases, standard MD simulations can become trapped in a local region of phase space, preventing reliable estimation of ensemble averages. 
This problem is particularly pronounced at lower temperatures, where activated processes such as octahedral tilting and organic-cation rotation occur infrequently.

To overcome these sampling limitations and ensure convergence of the ensemble averages required for TI, we employ replica exchange (REX) simulations.\cite{SUGITA1999_REX_NVT}  
REX involves running multiple parallel MD simulations at different temperatures, with Monte Carlo (MC) swap attempts between replicas.  
We restrict ourselves to temperature-based exchanges, although generalizations to varying pressures or Hamiltonians exist.
Swaps are attempted at intervals comparable to the MD correlation time to ensure sufficient decorrelation between configurations.\cite{Earl2005_REXreview}  

Specifically, at each swap attempt, the simulation evaluates the acceptance probability $P_{\text{accept}}$ for exchanging the temperatures of two NVT MD trajectories, based on their instantaneous energies $E_n$ and $E_m$, and temperatures $T_n$ and $T_m$:
\begin{equation}
	P_{\text{accept}} =  \min \left(1, \exp\left[\left( \frac{1}{k_\text{B}T_n} -  \frac{1}{k_\text{B}T_m} \right) \left( E_n - E_m \right) \right] \right).
	\label{AcceptanceProb}
\end{equation}
For NPT MD simulations conducted at the same pressure, the acceptance probability retains the same form as in Eq.\ \ref{AcceptanceProb}, with the energy replaced by the enthalpy.\cite{OKABE2001_REX_NPT}

An important advantage of the MC swaps is that they connect the phase space sampled by MD trajectories at different temperatures.  
The enhanced exploration of phase space at higher temperatures can thus assist lower-temperature simulations in escaping local minima.  
Through these swaps, energy barriers that are rarely crossed at low temperatures can be effectively overcome if they are frequently traversed at higher temperatures.  
As a result, REX simulations are employed to improve sampling for the TI contribution along the temperature integration path.  
Additional details regarding the REX implementation are provided in Sec.\ \ref{workflow}.

This sampling challenge also affects the TI correction between the harmonic and MLP PES, as it is evaluated at low temperature. 
Performing REX simulations for all $\lambda$ values between $0$ and $1$ would substantially increase the computational cost. 
Fortunately, this is not required: for small $\lambda$ values, sampling only needs to cover the vicinity of the harmonic minimum. 
For large $\lambda$ values, the potential is dominated by the MLP PES, and sampling at elevated temperatures becomes more appropriate, as the system can more readily overcome energy barriers. 
In contrast, at small $\lambda$, low-temperature sampling remains essential to avoid exploring unphysical regions of the PES that arise from the breakdown of the harmonic approximation.

To construct a continuous path between these two regimes, low $\lambda$ at low temperature and high $\lambda$ at high temperature, we introduce the intermediate PES, $U_\text{int}$.  
This PES provides a better approximation to the MLP PES than the harmonic PES, avoiding its breakdown at higher temperatures while retaining the harmonic minima as the dominant contribution at low temperature.
A practical choice is $U_\text{int} = U(0.7) = 0.3\, U_\text{harm} + 0.7\, U_\text{MLP}$, alternative intermediate PESs are discussed in Sec.\ \ref{workflow}.

Fig.\ \ref{Free_energy_drawing} schematically summarizes the required free energy contributions. 
We start from the ground state energy, $E_\text{GS}$ (contribution 1), and compute the static harmonic free energy, $F_\text{harm,stat}$ (contribution 2). 
Rather than directly applying a TI correction from the harmonic to the MLP PES, we first evaluate a TI correction from the harmonic to the intermediate PES at low temperature, $T_\text{low}$, where the harmonic approximation is more accurate (contribution 3). 
Next, a TI along a temperature path is performed for the intermediate PES using REX simulations to ensure convergence of the ensemble averages (contribution 4). 
Finally, a TI correction from the intermediate to the MLP PES is applied at high temperature, $T_\text{high}$, where sampling is more efficient (contribution 5).

The full temperature dependence of the Helmholtz free energy is then obtained by (see contributions 2 to 6 in Fig.\ \ref{Free_energy_drawing}):
\begin{equation}
	\begin{aligned}
		F_\text{MLP}(\textbf{h},T) & =\;
		F_\text{harm}(\textbf{h}, T_{\text{low}}) \\
		&+ \Delta_{\text{harm}\rightarrow \text{int}}^\text{TI} 
		F_{\text{int}}(\textbf{h},T_{\text{low}}) \\
		&+ \Delta_{T_\text{low}\rightarrow T_\text{high}}^\text{TI} 
		F_{\text{int}}(\textbf{h},T_\text{high}) \\
		&+ \Delta_{\text{int} \rightarrow \text{MLP}}^\text{TI} 
		F_\text{MLP}(\textbf{h},T_{\text{high}}) \\
		&+ \Delta_{T_\text{high} \rightarrow T}^\text{TI} 
		F_\text{MLP}(\textbf{h},T)
	\end{aligned}
	\label{eqTotalF}
\end{equation}

In principle, the Gibbs free energy can be computed using Eq.\ \ref{HelmToGibbs}, provided that the Helmholtz free energy is known for an arbitrary simulation cell (contribution 7). 
However, obtaining accurate probabilities for simulation cells that deviate significantly from the average cell in an NPT MD simulation at the target pressure and temperature is computationally unfeasible. 
Therefore, the Helmholtz free energy is evaluated at the temperature $T_\text{cell}$ (contribution 6) at which the average simulation cell (used for the Helmholtz free energy calculation) was determined.

To avoid repeating Helmholtz free energy calculations for the average cell at each temperature, we perform NPT REX simulations, from which the Gibbs free energy as a function of temperature can be obtained via TI along a temperature path (contribution 8). 
The full temperature dependence of the Gibbs free energy is then given by (see contributions 6 to 8 in Fig.\ \ref{Free_energy_drawing}):
\begin{equation}
	\begin{aligned}
	G_\text{MLP}(P,T) = & F_\text{MLP}(\textbf{h},T_\text{cell}) + \Delta_{\textbf{h}\rightarrow P} G_\text{MLP}(P,T_\text{cell}) + \\
	& \Delta_{T_\text{cell} \rightarrow T}^\text{TI} G_\text{MLP}(P,T),
\end{aligned}
	\label{eqTotalG}
\end{equation}
with $\textbf{h}$ the average simulation cell in a NPT MD simulation at pressure $P$ and temperature $T_\text{cell}$. 

Additionally, we perform NVE MD simulations to extract vibrational frequencies and compute the Helmholtz free energy (see $F_\text{harm,MD}$ in Fig.\ \ref{Free_energy_drawing}), as described in our previous work.\cite{Braeckevelt2022_CsPbI3} 
The free energies obtained from this method and our present approach are compared in Sec.\ \ref{FreeEnergyResults}.

This methodology assumes classical motion of the atomic nuclei. 
Quantum nuclear effects (QNE) can be incorporated through an alternative TI scheme,\cite{Kapil2019_JCTCv15} but this lies beyond the scope of the present work. 
In Sec.\ \ref{SI_quantumvsclassical} of the Supporting Information (SI), we investigate the difference between the harmonic free energy calculated using the classical partition function (see Eq.\ \ref{HarmFreeEnergy}) and the quantum mechanical partition function. 
As expected, the differences are small, since the high-frequency vibrational modes, which are most sensitive to quantum effects, are associated with the organic molecules, and these modes are similar across the three phases considered.

\section{Computational details}
\label{Computational_details}
\subsection{Density functional theory benchmark}
Because the ground state energy strongly influences the phase transition temperature, we assessed the accuracy of the energy and forces predicted by various XC functionals and dispersion corrections by benchmarking them against RPA+HF results. 
We used the same XC functionals, dispersion methods, and computational settings as in our previous study,\cite{Braeckevelt2022_CsPbI3} and extended the benchmark to include the organic MHPs: FAPbI$_3$ and MAPbI$_3$. 
In addition to energies, forces were also benchmarked, as they are directly used in training the MLP. 
To ensure meaningful force comparisons, we evaluated randomly selected structures from NPT MD simulations rather than optimized structures, which have near-zero forces by definition. 
RPA+HF forces were calculated in VASP using Green's function theory,\cite{Ramberger2017_RPAforces} and their convergence was verified, as discussed in Sec.\ \ref{SI_Convergence_tests} of the SI.

\subsection{Training of the machine learning potentials}
\label{TrainingMLP}
\subsubsection{\textit{Ab initio} data generation}

To generate the dataset, we first performed NVT \textit{ab initio} MD simulations at 800 K using CP2K.\cite{CP2K1,CP2K2} 
The simulations employed the PBE XC functional with D3(BJ) dispersion corrections.\cite{PBE,D3BJ} 
CP2K utilizes a mixed Gaussian and plane-wave basis, and we used GTH pseudopotentials in combination with TZVP-MOLOPT-SR-GTH basis sets. 
The plane-wave cutoff and relative cutoff were set to 400 Ry and 40 Ry, respectively. 
Each simulation box contained 64 formula units (fu), and only the $\Gamma$-point was used for Brillouin zone sampling.

For each material and each phase, MD simulations were performed at 22 different volumes to generate training structures for the MLP, and at 5 additional volumes to generate validation and test structures. 
A volume range of approximately $\pm$10\% around the average volume at 300 K was uniformly sampled.
Initial configurations were extracted from an NPT MD simulation at 800 K and 0.1 MPa; one structure was selected every 100 ps and isotropically scaled to the desired volume. 
For FAPbI$_3$ and MAPbI$_3$, the organic molecules were additionally subjected to random rotations to reduce correlations between the NVT MD simulations and to ensure a more uniform sampling of the phase space. 
During the NVT and NPT simulations with CP2K, a Nosé-Hoover thermostat with a time constant of 100 fs and three thermostat beads was used to control the temperature.\cite{NoseHooverThermostat1,NoseHooverThermostat2,NoseHooverThermostat3} 
For the NPT simulations, the pressure was maintained using a Martyna-Tuckerman-Klein (MTK) barostat with a time constant of 500 fs.\cite{MTK}

We performed 1000 steps for each NVT MD simulation, using a time step of 2 fs for CsPbI$_3$ and 0.5 fs for FAPbI$_3$ and MAPbI$_3$. 
From each MD simulation in the training dataset, 6 structures were uniformly selected. 
For the validation and test datasets, 2 structures were selected from each of the five MD simulations at the additional volumes for validation, and 2 different structures were selected for testing.
Three separate MLPs were trained, one for each material. 
By combining the structures from all phases of a given material, we constructed datasets consisting of 396 training structures, 30 validation structures, and 30 test structures.

The initial dataset was sufficient to train MLPs that accurately reproduced the \textit{ab initio} PES sampled at temperatures up to 600 K. 
However, our TI approach also requires sampling the harmonic (Hessian) PES at lower temperatures, specifically 150 K. 
While the MLPs were generally accurate for most structures generated during MD on the Hessian PES, a few outlier structures exhibited large prediction errors, which strongly affected the ensemble averages required for the free energy calculation (see Eq.\ \ref{lambdaTI}). 
To address this, we developed new MLPs for each material by augmenting the original training dataset with 48 additional structures sampled from NVT MD simulations on the Hessian PES at 150, 245, 383, and 600 K for all three phases. 
For each phase and temperature, seven structures were added to the training set and one to the validation set. 
These updated MLPs were then used to compute the free energies, as they accurately described both the true and Hessian PES across the relevant temperature range.

Before training the MLPs, all structures in the dataset were recalculated using VASP\cite{VASP1,VASP2}, which yields low net forces,\cite{Kuryla2025_netforces} with the PBE+D3(BJ) XC functional.\cite{PBE,D3BJ}
The following PAW potentials\cite{PAW} were employed: H (1$s^1$), C (2s$^2$2p$^2$), N (2s$^2$2p$^3$), Cs\_sv (5s$^2$5p$^6$6s$^1$), Pb\_d (5d$^{10}$6s$^2$6p$^2$), and I (5s$^2$5p$^5$). 
A $\Gamma$-centered $2 \times 2 \times 2$ k-point grid and a plane-wave cutoff energy of 500 eV were used. 
This recalculation step was necessary because the CP2K-predicted energy differences between phases differed from the VASP values by several kJ/mol. 
Moreover, the phase energy differences obtained with PBE+D3(BJ) in VASP show good agreement with the RPA+HF reference values, as discussed in Sec.\ \ref{DFTbenchmark}.

The CP2K MD input files and the VASP single-point calculation input files are available at https://doi.org/10.5281/zenodo.18108633.

\subsubsection{Architecture of the machine learning potentials}
The MACE architecture with a cut-off radius of 6 \AA{} was used to train the MLP. 
We employed the psiflow software package (version 1.0.0),\cite{Vandenhaute2023_psiflow} and adapted several settings of the MACE architecture as implemented in psiflow; these are listed in Tab.\ \ref{settingsMACE}.
The input files used for MLP training are available at https://doi.org/10.5281/zenodo.18108633.
This relatively small network could be evaluated on 1 cpu core. 
The evaluation time per structure for 8 and 64 fu simulation cells of CsPbI$_3$ was 32 and 183 ms, respectively. 
For the organic MHPs, the corresponding evaluation times were 86 and 685 ms.
These fast evaluation times are essential, as more than 200 million MD steps are required to compute the Gibbs free energy of a single system (see next subsection).

\begin{table}[h]
	\small
	\caption{\ Non-default settings of the applied MACE architecture used for MLP training}
	\label{settingsMACE}
	\begin{tabular*}{0.48\textwidth}{@{\extracolsep{\fill}}lll}
		\hline
		\textbf{Setting} & \textbf{Value} \\
		\hline
		\texttt{r\_max} & 6.0 \\
		\texttt{max\_ell} & 2 \\
		\texttt{num\_channels} & 8 \\
		\texttt{lr} & 0.02 \\
		\texttt{max\_num\_epochs} & 300 \\
		\texttt{ema} & True \\
		\hline
	\end{tabular*}
\end{table}

For the MLPs trained on the CsPbI$_3$, FAPbI$_3$, and MAPbI$_3$ datasets (including the Hessian structures), the test set energy errors were 0.4, 0.3, and 0.2 meV/atom, respectively, while the corresponding force errors were 25.6, 28.7, and 24.3 meV/\AA. 
These root mean square errors (RMSEs) are significantly smaller than the standard deviations of the respective test sets, which were 18.4, 6.8, and 5.8 meV/atom for the energies and 358.5, 1175.8, and 1068.3 meV/\AA\ for the forces.

\subsection{Workflow}
\label{workflow}
\begin{figure}[h]
	\centering
	\includegraphics[width=1.0\linewidth]{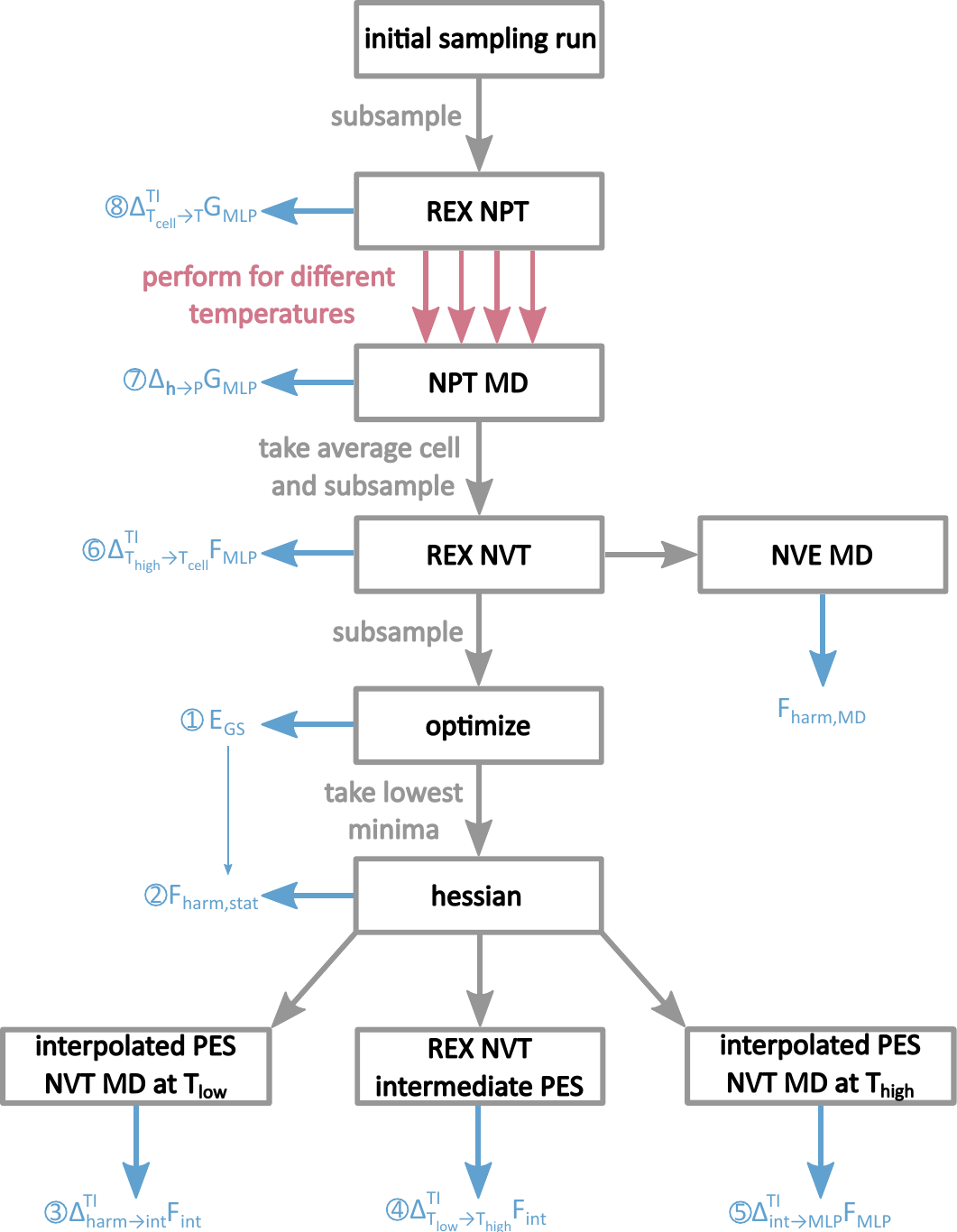}
	\caption{Flowchart to derive the Gibbs free energy of a phase.
		The numbers next to the contributions indicate the order in which they are theoretically added, as depicted in Fig.\ \ref{Free_energy_drawing}.
		As explained in our previous work,\cite{Braeckevelt2022_CsPbI3} the harmonic free energy at a specific temperature can also be determined via an additional NVE MD simulation.}
	\label{SchematicOverviewMethod}
\end{figure}

The PES of MHPs contains numerous local minima, which hampers the convergence of the ensemble averages required for free energy calculations. 
To obtain converged values, we perform multiple MD simulations in parallel and employ replica exchange (REX) simulations. 
To accelerate convergence, it is important that parallel simulations are initialized from distinct random structures that are still thermodynamically relevant at the conditions of interest. 
Each contribution depicted in Fig.\ \ref{Free_energy_drawing} requires a separate calculation, and the sampling from one step is used to generate the initial structures for the next. 
The general workflow is illustrated in Fig.\ \ref{SchematicOverviewMethod}. 
All calculations in this workflow are performed using the trained MLPs, as discussed in Sec.\ \ref{TrainingMLP}.

First, we define the temperature range of interest. 
The MD simulation at the highest temperature should be able to easily escape local minima and efficiently explore the entire PES of the phases of interest. 
The lowest temperature is ideally chosen to be as low as possible, but is constrained by the number of replicas and the acceptance probability of the Monte Carlo exchanges in the REX simulations. 
For all MHPs investigated, we use 32 temperatures ranging from 150 K to 600 K. 
The temperatures follow a geometric progression ($\frac{T_{i+1}}{T_i} = \text{constant}$) to ensure that all replica swaps have a similar acceptance ratio.\cite{Earl2005_REXreview}

We begin with a short NPT MD simulation at the highest temperature (600 K) and 0.1 MPa to generate 32 distinct initial structures for the REX NPT simulations. 
The REX NPT simulation couples 32 MD simulations at different temperatures under constant pressure (0.1 MPa). 
From this simulation, the temperature-dependent free energy contribution, $\Delta_{T_\text{cell} \rightarrow T}^\text{TI} G_\text{MLP} (P,T)$, can be computed.

In principle, the $\Delta_{\mathbf{h} \rightarrow P} G_\text{MLP}(P,T_\text{cell})$ correction could be determined by selecting one temperature, calculating the average simulation cell, and evaluating the corresponding $\Delta_{\textbf{h}\rightarrow{} P} G_\text{MLP} (P,T_\text{cell})$ correction.
However, accurately sampling $\rho(\mathbf{h}|P,T)$ is challenging because the cell shape matrix $\mathbf{h}$ is six-dimensional. 
To reduce complexity, we instead approximate it using the volume probability distribution $\rho(V|P,T)$. 
Moreover, at low temperatures, $\mathbf{h}$ may exhibit multiple minima due to different octahedral tilt patterns. 
To obtain a meaningful average structure, a separate NPT MD simulation is performed at each temperature, as the cell parameters tend to remain in a local minimum at low temperature. 
The average cell from this simulation is used, and the probability of the corresponding volume, $\rho(V|P,T)$, is used to compute $\Delta_{\mathbf{h} \rightarrow P} G_\text{MLP}(P,T_\text{cell})$.
To assess the impact of the approximations made, this correction (and the associated Helmholtz free energy $F_\text{MLP}(\mathbf{h},T_\text{cell})$) is evaluated at eight different temperatures: 150, 187, 234, 293, 350, 419, 501, and 600 K.

Due to the rough PES with numerous local minima, we first explore the PES and subsequently perform multiple geometry optimizations. 
To generate a relevant set of structures for optimization, MD simulations are preferably performed at the lowest temperature. 
However, to escape local minima and ensure broad sampling of the PES, MD simulations at the highest temperature are necessary. 
To achieve both objectives, we perform REX NVT simulations and extract 4000 structures from the replica at the lowest temperature for geometry optimization.
Moreover, the REX NVT simulations enable the calculation of the free energy contribution $\Delta_{T_\text{high} \rightarrow T_\text{cell}}^\text{TI} F_\text{MLP}(\mathbf{h},T_\text{cell})$.
From the set of optimized structures, the one with the lowest energy is selected, and its Hessian is computed. 
This Hessian is then used to determine the harmonic free energy as a function of temperature.

For CsPbI$_3$, we use an intermediate PES defined as $0.3 U_\text{harm} + 0.7 U_\text{MLP}$. For FAPbI$_3$ and MAPbI$_3$, the intermediate PES corresponds to the MLP PES augmented with a bias potential. 
This bias potential constrains the orientation of the organic molecules, effectively fixing them during simulations on the intermediate PES. 
As a result, only the minima of the harmonic PES are sampled, which circumvents convergence issues that typically arise at low temperatures. 
Details about the construction and implementation of the bias potential are provided in Sec.\ \ref{SI_biaspotential} of the SI.

REX NVT simulations at various temperatures are performed on the intermediate PES to compute the thermodynamic integration (TI) correction $\Delta_{T_\text{low} \rightarrow T_\text{high}}^\text{TI} F_\text{int}(\textbf{h},T_\text{high})$. 
In addition, interpolated PESs between the Hessian PES and the intermediate PES are constructed to perform MD simulations at the lowest temperature, yielding the TI correction $\Delta_{\text{harm} \rightarrow \text{int}}^\text{TI} F_\text{int}(\textbf{h},T_\text{low})$. 
Similarly, MD simulations between the intermediate and MLP PES are used to compute the correction $\Delta_{\text{int} \rightarrow \text{MLP}}^\text{TI} F_\text{MLP}(\textbf{h},T_\text{high})$.
For both TI corrections, 14 intermediate PESs were constructed using Eq.\ \ref{eq_intPES}, with the following $\lambda$ values: 0.0, 0.01, 0.02, 0.05, 0.1, 0.2, 0.4, 0.6, 0.8, 0.9, 0.95, 0.98, 0.99, and 1.0.

To investigate the effect of supercell size, shorter MD simulations were performed using simulation cells containing 64 fu instead of 8.  
Because the phase space is considerably larger for the bigger cell, the final structure from the 8 fu MD simulations was used as the initial structure for the $2 \times 2 \times 2$ supercell to reduce equilibration time.  
The reduce the overall computational cost, the Helmholtz free energy, $F_\text{MLP}(\mathbf{h}, T_\text{cell})$, for 64 fu was calculated using the average cell at only three temperatures: 234 K, 350 K, and 501 K.

All structural optimizations were carried out with ASE\cite{HjorthLarsen2017} using the PreconLBFGS optimizer\cite{Packwood2016}.  
MD simulations and Hessian calculations were performed with Yaff.\cite{yaff}  
Temperature and pressure were controlled with a Langevin thermostat and barostat,\cite{Langevin1,Langevin2} using a timestep of 2 fs for CsPbI$_3$ and 0.5 fs for FAPbI$_3$ and MAPbI$_3$.  
Multiple parallel MD simulations were performed for each contribution, resulting in nearly 220 million MD steps required to compute the Gibbs free energy of a single material at a given system size.
The exact number of simulation steps for each contribution is provided in the simulation scripts available at https://doi.org/10.5281/zenodo.18108633.

\section{Results}
\subsection{Benchmarking XC functionals with RPA+HF}
\label{DFTbenchmark}
We benchmarked the energies and forces predicted by various XC functionals and dispersion corrections for CsPbI$_3$, FAPbI$_3$, and MAPbI$_3$. 
For each material and each of the three phases, five representative structures were randomly selected from NPT MD simulations at 600 K and 0.1 MPa using a simulation cell containing four fu, resulting in a total of 45 structures for the energy evaluation. 
Given the high computational cost of RPA+HF force calculations, a subset of two structures per phase and material (\textit{i.e.}, 18 in total) was selected for force benchmarking.

\begin{figure*}
	\centering
	\includegraphics[width=1.0\linewidth]{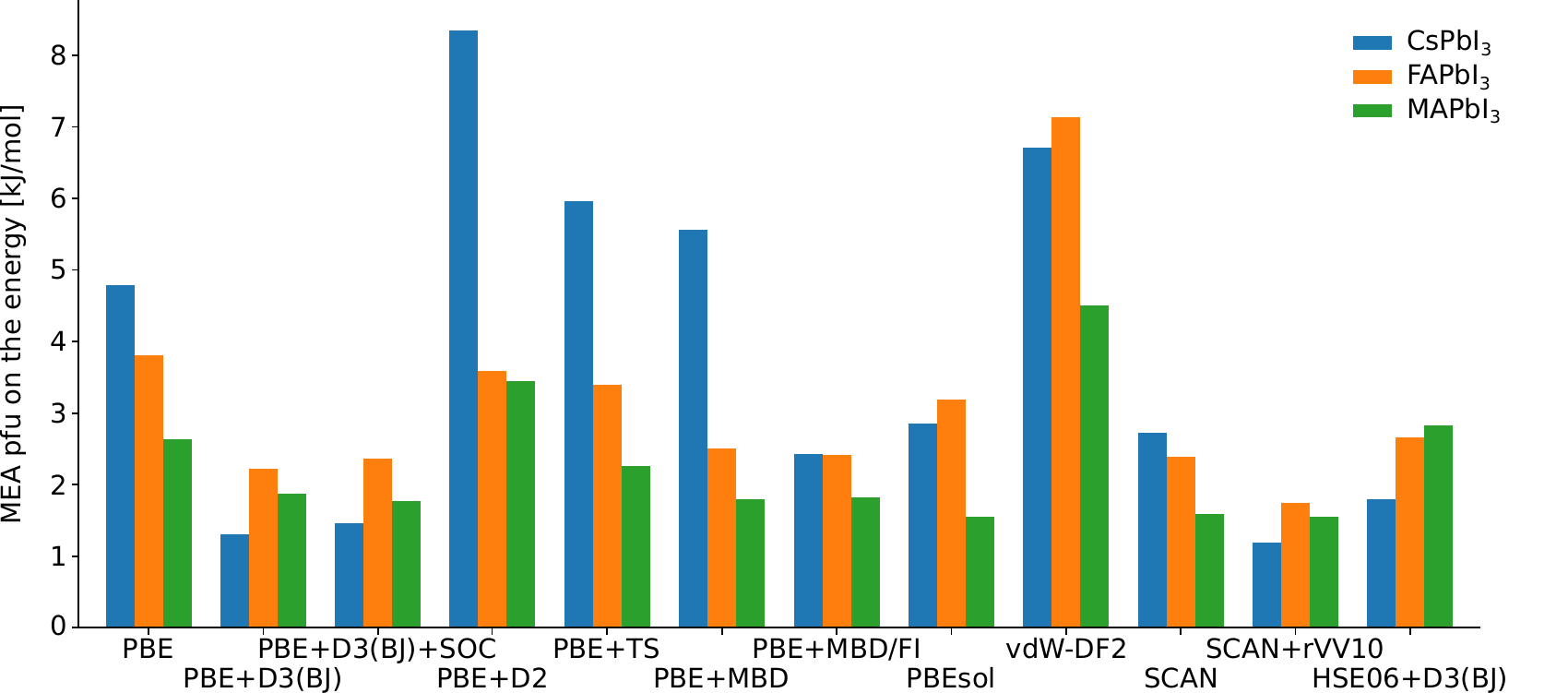}
	\caption{Mean absolute error of the DFT energies using different XC functionals and dispersion corrections with respect to RPA+HF.}
	\label{Error_energy_random}
\end{figure*}

The mean absolute errors (MAEs) of the DFT energies with respect to the RPA+HF reference are shown in Fig.\ \ref{Error_energy_random}. 
With the exception of vdW-DF2 and a few outliers for CsPbI$_3$, most XC functionals and dispersion schemes yield comparable accuracy. 
Among these, SCAN+rVV10 achieves the lowest MAE of 1.5 kJ/mol per formula unit (pfu), demonstrating the best agreement with RPA+HF.\cite{Xue2021_DFTbenchmark}
PBE+D3(BJ) also performs well, with an average MAE of 1.8 kJ/mol pfu.

A similar benchmark performed on optimized structures is presented in Sec.\ \ref{SI_LOT_comp_GSE} of the SI and yields consistent trends. 
Furthermore, for several optimized structures used in the thermodynamic integration procedure, we compared the MLP-predicted energies with PBE+D3(BJ) values, finding deviations smaller than 1 kJ/mol pfu. 

\begin{figure*}
	\centering
	\includegraphics[width=1.0\linewidth]{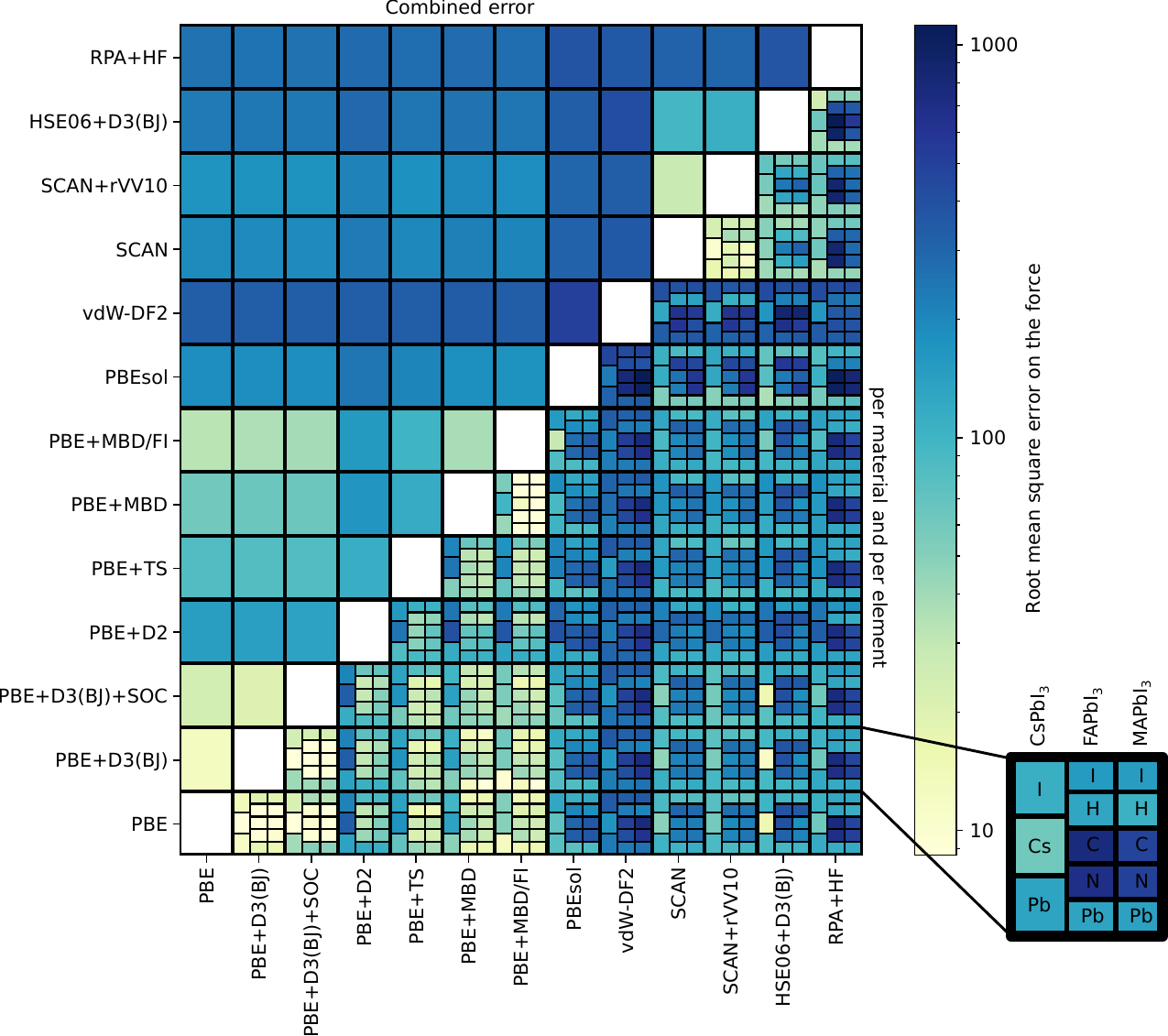}
	\caption{
		Root mean square error of the atomic forces between different XC functionals, dispersion methods, and RPA+HF calculations. 
		The top left shows the total RMSE across all atoms from two benchmark structures per phase and material. 
		The bottom right breaks down the RMSE by material and atomic species.
	}
	\label{Error_forces_summary}
\end{figure*}

The RMSEs of the forces between the different levels of theory are shown in Fig.\ \ref{Error_forces_summary}. 
Variations in the dispersion correction have only a minor effect on the computed forces. 
For the PBE XC functional, the largest deviations occur when using the D2 dispersion correction, primarily due to inaccuracies in the forces on Cs atoms. 
In contrast, adding D3(BJ), SOC, or MBD/FI dispersion corrections to PBE does not significantly alter the force predictions compared to PBE without dispersion. 

Similarly, SCAN, SCAN+rVV10, and HSE06+D3(BJ) show good mutual agreement, indicating internal consistency among these functionals. 
In contrast, vdW-DF2 stands out as a clear outlier, yielding large RMSEs in the forces when compared to all other methods. 
While PBEsol exhibits better agreement than vdW-DF2, it still shows RMSEs exceeding 170 meV/\AA{} with respect to the other functionals, indicating notable discrepancies.

No XC functional or dispersion method shows significantly better agreement with RPA+HF for the atomic forces. 
Among the tested methods, PBE+D3(BJ) yields the lowest overall RMSE of 260 meV/\AA, while PBEsol results in the highest RMSE of 381 meV/\AA. 
However, a breakdown of the RMSE by atomic species reveals considerable variation. 
Specifically, the errors on the C and N atoms are substantially larger than those on Cs, Pb, and I atoms. 
This discrepancy may be attributed to the delocalized electronic charge distribution on the C and N atoms in the FA and MA molecules, which standard DFT functionals often fail to accurately describe.\cite{Cohen2012,Broberg2023} 
Furthermore, the RMSE is slightly higher for the C and N atoms of FA compared to those of MA, possibly because the electronic charge in FA is more delocalized than in MA.

While the total RMSE of the atomic forces with respect to RPA+HF is comparable across GGA, mGGA, and hybrid XC functionals, significant differences arise when partitioning the error between organic and inorganic atoms.
The mGGA and hybrid functionals yield RMSE values below 85 meV/\AA\ for the inorganic atoms, whereas the RMSE on the organic atoms can exceed 850 meV/\AA. 
In contrast, the PBE+D3(BJ) functional results in higher RMSE values for the inorganic atoms, up to 140 meV/\AA. 
However, it provides a more accurate description of the forces on the organic species.

In general, we expect that the atomic dynamics predicted by most DFT methods, when compared to RPA+HF, primarily differ in the internal vibrational modes of the FA and MA molecules.
These modes are present in all three phases.
Consequently, the large force errors observed on the organic atoms are not expected to significantly impact the free energy differences between phases, due to a cancellation of errors. 

We opted to train the MLPs on PBE+D3(BJ) energies and forces computed with VASP\cite{VASP1,VASP2}, as this functional yields one of the best agreements with RPA+HF while being computationally less demanding than alternatives such as SCAN+rVV10.
Furthermore, the PBE functional is known to produce less noisy forces compared to many other XC functionals.\cite{Sitkiewicz2022}

\subsection{Free energy as a function of temperature}
\label{FreeEnergyResults}
\begin{figure*}
	\centering
	\includegraphics[width=1.0\linewidth]{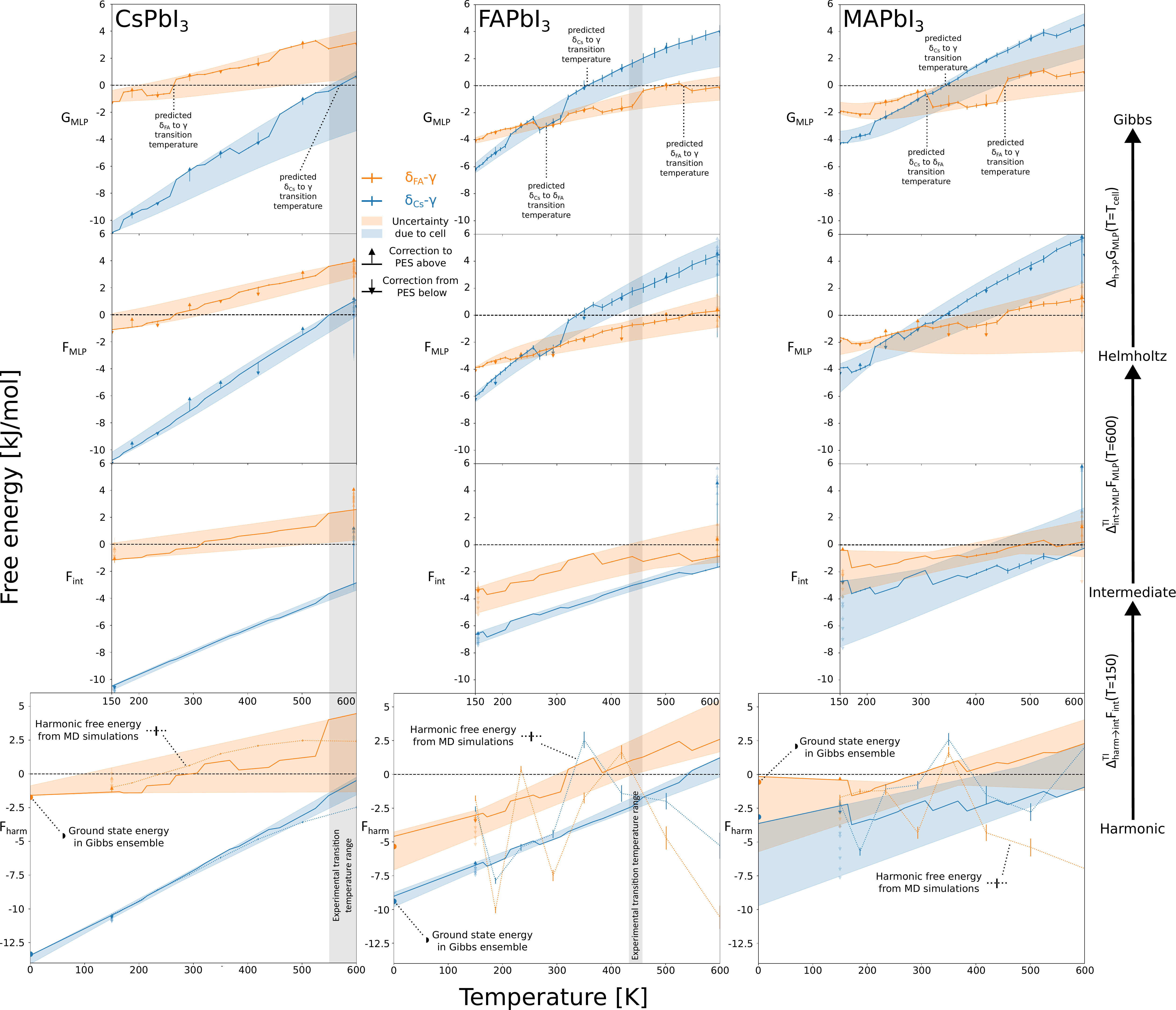}
	\caption{
		Free energy difference of the $\delta_\text{Cs}$ (blue) and $\delta_\text{FA}$ (orange) phases relative to the $\gamma$ phase as a function of temperature, computed using different PES: 
		harmonic (via NVE MD simulations, dashed lines, and via Hessian calculations, solid lines with shaded areas), intermediate, and MLP. 
		The Helmholtz and Gibbs free energies for the MLP PES are obtained from MD simulations in the NVT and NPT ensembles, respectively. 
		All results correspond to simulation cells containing eight fu of CsPbI$_3$, FAPbI$_3$, and MAPbI$_3$. 
		For the NVT simulations, eight independent simulation cells were used; the shaded areas represent the range between the minimum and maximum values. 
		The solid lines depict the free energy computed for the average NPT MD simulation cell at the temperature closest to the corresponding x-axis value on a logarithmic scale.
		The arrows originating from the solid lines represent corrections to the free energy plotted above, while the arrows pointing toward the solid lines represent corrections arising from the free energy plotted below.}
	\label{Free_energy_results_8fu}
\end{figure*}

By performing the simulations outlined in Fig.\ \ref{SchematicOverviewMethod} with the trained MLP, the four temperature-dependent free energies shown in Fig.\ \ref{Free_energy_drawing} can be computed.
As detailed in Sec.\ \ref{SI_FEC_analysis} of the SI, each simulation step contributes to the total free energy and can be analyzed independently. 
Fig.\ \ref{Free_energy_results_8fu} presents the four free energy curves as a function of temperature for CsPbI$_3$, FAPbI$_3$, and MAPbI$_3$.
For all MHPs, the $\delta$ phases are stabilized by enthalpy, whereas the $\gamma$ phase is stabilized by entropy.\cite{Liu2022_reversiblephasetransitions}

The bottom panel for each material shows the harmonic free energy. 
This quantity can be computed using two different methods: 
from the vibrational frequencies obtained from the Hessian of the ground state structure (solid lines), 
and from the velocities in NVE MD simulations (dotted lines).\cite{Braeckevelt2022_CsPbI3}
For CsPbI$_3$, both approaches yield consistent results, indicating good agreement. 
However, for FAPbI$_3$ and MAPbI$_3$, the harmonic free energy derived from the velocities of the NVE MD simulations is less accurate and fails to converge. 
As discussed in Sec.\ \ref{SI_histogram_opt_energies_section} of the SI, this discrepancy arises due to the presence of numerous inequivalent local minima introduced by the organic molecules. 
These contribute to poor convergence of the frequencies and, consequently, the harmonic free energy when derived from MD trajectories.

The intermediate PES is designed to incorporate anharmonic effects while remaining constrained to the region near the global energy minimum. 
Comparison of the bottom panel with the panel above shows that this correction is small and tends to stabilize the $\delta$ phases relative to the $\gamma$ phase at elevated temperatures.

The correction from the intermediate to the MLP PES includes the contribution from configurational entropy.
Comparison of the second and third panels shows that this correction slightly destabilizes the $\delta_\text{FA}$ phase and significantly destabilizes the $\delta_\text{Cs}$ phase.
These results suggest that the $\gamma$ phase has access to a larger number of local minima than the $\delta_\text{FA}$ phase, which in turn has more than the $\delta_\text{Cs}$ phase. 
Interestingly, for MAPbI$_3$, the correction destabilizes the $\gamma$ phase at low temperatures, indicating that some local minima become accessible only at elevated temperatures (see also Sec.\ \ref{Asite_effect}).

To calculate the Helmholtz free energy, we used eight different simulation cells, resulting in a spread indicated by the shaded area in Fig.\ \ref{Free_energy_results_8fu}. 
This best estimate, shown as solid lines, corresponds to the average NPT MD simulation cell at the temperature closest (on a logarithmic scale) to the one considered. 
Using Eq.\ \ref{HelmToGibbs}, we applied a correction to convert the Helmholtz free energies into Gibbs free energies for all eight cells (top panel). 
Ideally, this correction should collapse the spread, since all Gibbs free energies at a given $(P,T)$ should be equal. 
However, the spread remains comparable, likely because the correction is based on the volume distribution $\rho(V|P,T)$ rather than the full six-dimensional cell distribution $\rho(\mathbf{h}|P,T)$. 

The experimental transition temperature of CsPbI$_3$ from the $\delta_\text{Cs}$ to the $\gamma$ phase ranges from 548 K to 599 K, in good agreement with the prediction from the TI approach. 
For FAPbI$_3$, the experimental transition from the $\delta_\text{FA}$ to the $\gamma$ phase occurs between 433 K and 458 K, with the TI method slightly overestimating this value.\cite{Dastidar2017,Wang2019Ttrans,An2021,Han2016,Jeon2015} 
In the case of MAPbI$_3$, the $\gamma$ phase is experimentally stable at room temperature, which is not reproduced by the TI predictions. 
Nevertheless, the free energy differences between the three phases at room temperature are very small, suggesting that only a modest correction would be required to align the predictions with experimental observations.

Based on the limited experimental data for the temperatures at which the free energy of two phases are the same, we conclude that the TI approach predicts small free energy differences at the experimental transition temperatures. 
However, there is some degree of error cancellation, as the free energy difference relative to the $\gamma$ phase decreases when using a supercell containing 64 fu (see Sec.\ \ref{SI_free_energy_64fu} of the SI). 
This results in free energy differences at the experimental transition temperatures that are approximately 2 kJ/mol pfu too low, which may stem from inaccuracies in the PBE+D3(BJ) XC functional. 
Additional sources of error could include insufficient sampling of the phase space, convergence issues, or the neglect of quantum nuclear effects. 
Moreover, experimental factors such as the presence of defects may also influence the observed transition temperatures.

\subsection{A-site effect on the free energy}
\label{Asite_effect}

\begin{figure*}
	\centering
	\includegraphics[width=1.0\linewidth]{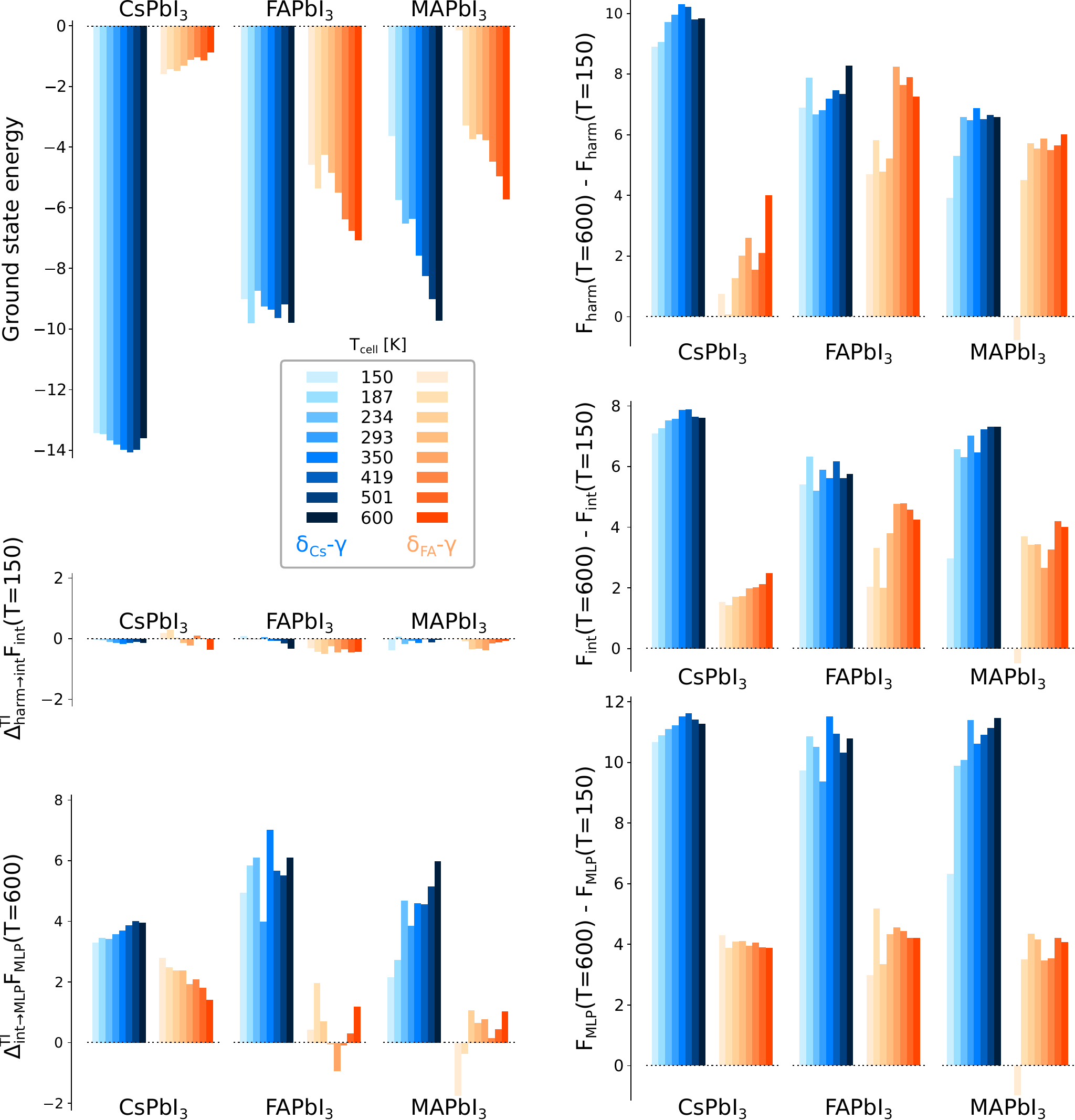}
	\caption{
		Contributions to the total Helmholtz free energies (in kJ/mol pfu) calculated for average simulation cells at eight different temperatures $T_\text{cell}$, as described in Eq\ \ref{eqTotalF}. 
		The energy differences of the $\delta$ phases relative to the $\gamma$ phase are shown; positive values indicate stabilization of the $\gamma$ phase. 
		The left panel presents the ground-state energies and the $\lambda$-TI corrections for the intermediate and MLP PES. 
		The right panel illustrates the temperature-dependent free energy changes for the harmonic, intermediate, and MLP PES.
	}
	\label{Free_energy_contributions_8fu}
\end{figure*}

To investigate the effect of the A-site cation on the free energy, we analyze six contributions to the Helmholtz free energy. 
Fig.\ \ref{Free_energy_contributions_8fu} presents the ground-state energy of the global minima, the two $\lambda$-TI corrections, and the difference in Helmholtz free energy between 600 K and 150 K for the three different PES.

The top-left panel shows the ground-state energy and is strongly influenced by the A-site cation, with a pronounced difference between CsPbI$_3$ and the organic MHPs.
The $\delta_\text{Cs}$ phase is significantly more stable in CsPbI$_3$, whereas the $\delta_\text{FA}$ phase is favored in the organic MHPs, with FAPbI$_3$ exhibiting slightly greater stability in this phase compared to MAPbI$_3$.
The cell has a notable impact, particularly for the organic MHPs, where the average cells at elevated temperatures tend to destabilize the $\gamma$ phase.

As shown in the top-right panel of Fig.\ \ref{Free_energy_contributions_8fu}, the harmonic free energy differences show the opposite trend, stabilizing the $\gamma$ phase at elevated temperatures.
For the $\delta_\text{Cs}$ phase, this stabilization is more pronounced in CsPbI$_3$, while for the $\delta_\text{FA}$ phase, it is stronger in the organic MHPs.
TI corrections to the intermediate PES and their temperature dependence shown by the middle panels in Fig.\ \ref{Free_energy_contributions_8fu}, introduce only minor changes to the harmonic free energy across all materials, slightly reducing cell size dependence and destabilizing the $\gamma$ phase.

The correction to the MLP PES is reported in the bottom-left panel of Fig.\ \ref{Free_energy_contributions_8fu} and includes configurational entropy, which has a significant influence on the total free energy. 
All three materials share the same PbI framework, whose flexibility is most easily analyzed in CsPbI$_3$. 
For instance, when considering the $\gamma$ phase of CsPbI$_3$ using a simulation cell with eight fu at 600 K, all but one of the 4000 sampled structures optimize to the same energy, as shown in Sec.\ \ref{SI_histogram_opt_energies_section} of the SI. 
However, this does not imply that they converge to the same point in phase space. 
Due to the various possible octahedral tilting directions and axes, multiple distinct local minima can exist with nearly identical energies.\cite{Bechtel2018} 
A comparison of the distance matrices of the atomic positions across all optimized structures revealed 25 unique configurations, indicating that the $\gamma$ phase contains 24 equivalent minima. 
In contrast, only one unique configuration was identified for the $\delta_\text{Cs}$ phase, and six configurations with approximately the same energy were found for the $\delta_\text{FA}$ phase.

The contribution of configurational entropy to the free energy can be estimated by taking the natural logarithm of the number of distinct minima and multiplying by $k_\text{B}T$. 
At 600 K, this results in a destabilization of the $\delta_\text{Cs}$ phase by approximately 1.2 kJ/mol pfu relative to the $\delta_\text{FA}$ phase, and 2 kJ/mol pfu relative to the $\gamma$ phase. 
These values are slightly lower than $\Delta_{\text{int} \rightarrow{} \text{MLP}}^\text{TI} F_\text{MLP} (\mathbf{h}_{600},T=600)$ shown in Fig.\ \ref{Free_energy_contributions_8fu}, as the latter also accounts for contributions from higher-energy minima and anharmonic effects.

The configurational entropy is influenced by the size of the supercell.
For the configurational entropy to scale linearly with system size, the number of accessible minima must scale exponentially. 
This condition is not met for the $\gamma$ phase of CsPbI$_3$, as octahedral tilts in one part of the structure constrain the tilts in neighboring regions. 
While larger supercells can accommodate more local minima through irregular tilting patterns, these configurations generally have higher energies. 
As a result, normalizing the configurational entropy contribution pfu leads to a relative destabilization of the $\gamma$ phase when increasing the system size, see Sec.\ \ref{SI_free_energy_64fu} of the SI. 
In contrast, this destabilizing effect is not observed for the $\delta_\text{FA}$ phase compared to the $\delta_\text{Cs}$ phase, indicating that the number of energetically relevant minima in the $\delta_\text{FA}$ phase scales exponentially with system size.

The rotational freedom of the organic molecules generates a significantly larger number of local minima, as shown in Sec.\ \ref{SI_histogram_opt_energies_section} of the SI. 
By comparing the $\Delta_{\text{int} \rightarrow \text{MLP}}^\text{TI} F_\text{MLP}(\textbf{h}_{600},T=600)$ contributions for FAPbI$_3$ and CsPbI$_3$ (\textit{i.e.}, the dark blue bars of both materials and the dark red bars in the bottom-left panel of Fig.\ \ref{Free_energy_contributions_8fu}), we infer that the FA molecule exhibits reduced rotational freedom in the $\delta_\text{Cs}$ phase and enhanced rotational freedom in the $\delta_\text{FA}$ phase relative to the $\gamma$ phase.
A similar trend is observed for the MA molecule, although the cell dependence is significantly more pronounced for MAPbI$_3$.

\begin{figure*}
	\centering
	\includegraphics[width=1.0\linewidth]{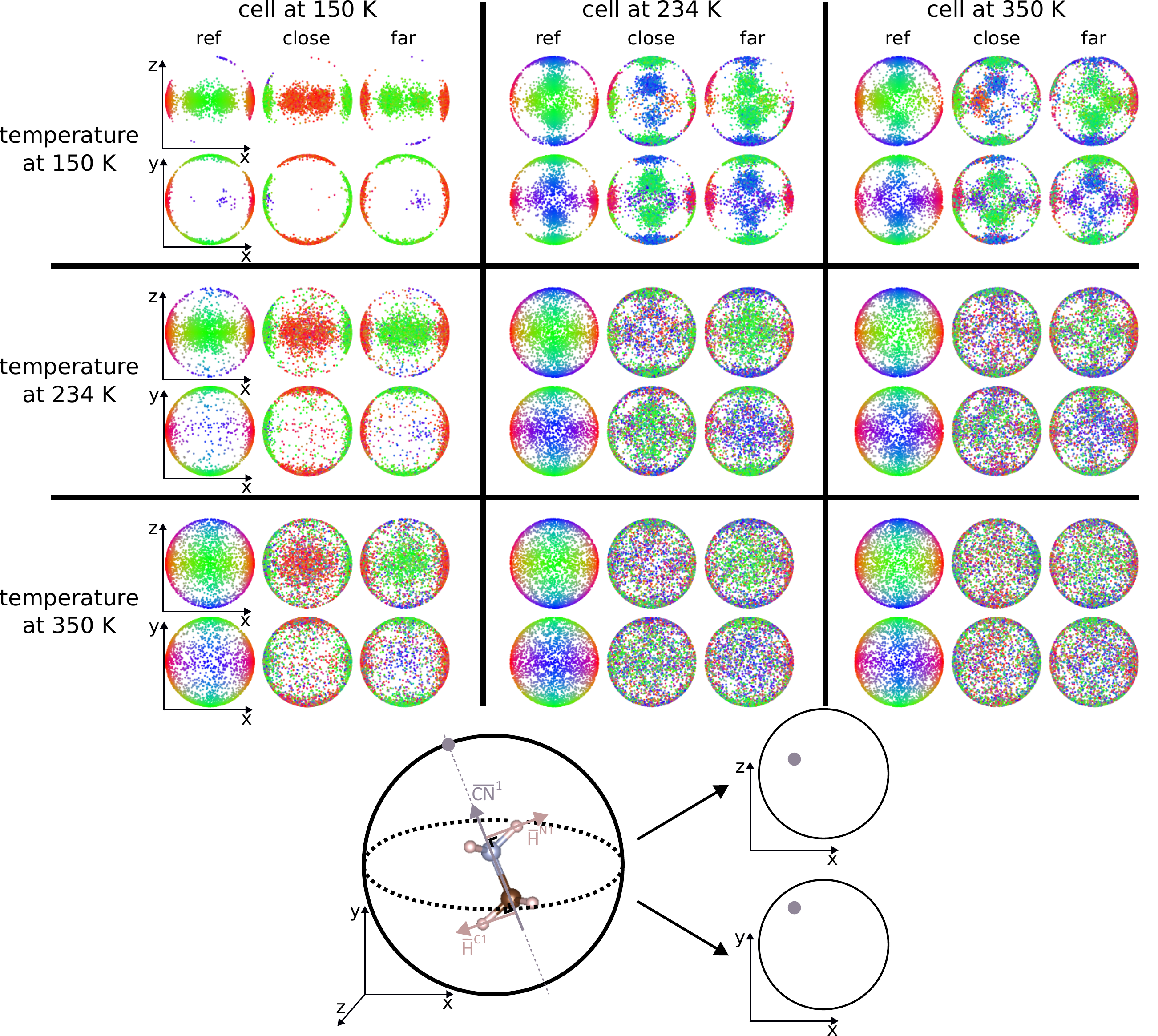}
	\caption{
		Orientation distributions of three MA molecules in the $\gamma$ phase of MAPbI$_3$ during NVT simulations on the MLP PES at the specified temperature and simulation cell. 
		The orientation of each molecule is represented by the unit vector $\overrightarrow{\text{CN}}$, defined from the carbon to the nitrogen atom. 
		The endpoints of these vectors are projected onto the $xz$ and $xy$ planes, as illustrated at the bottom of the figure. 
		For each simulation, three MA molecules are selected: one is randomly chosen as the reference, and the other two are the nearest and furthest molecules from the reference. 
		The RGB color of each dot is determined by the $x$, $y$, and $z$ components of the $\overrightarrow{\text{CN}}$ vector of the reference MA molecule in the simulation cell.
	}
	\label{Orientation_plots_MAPbI3_cell}
\end{figure*}

To understand the effect of the cell, we investigate the orientational freedom of the organic molecules by tracking the vector defined by the carbon and nitrogen atoms of a MA molecule. 
Fig.\ \ref{Orientation_plots_MAPbI3_cell} presents the orientation distributions for three MA molecules in the $\gamma$ phase of MAPbI$_3$ during NVT MD simulations at various temperatures and for different simulation cell sizes. 

Instead of the spread of the orientation distributions, we investigate correlations between the orientations of different molecules, as these can strongly affect the number of energetically relevant minima and, consequently, the configurational entropy. 
If the orientations are perfectly correlated, the number of minima is determined by a single molecule, and the configurational entropy does not increase with system size. 
Conversely, if molecular orientations are independent, adding a molecule with two possible orientations doubles the number of minima, and the configurational entropy grows accordingly.

To visualize correlations between different molecules, the RGB color of each dot in Fig.\ \ref{Orientation_plots_MAPbI3_cell} is determined by the $x$, $y$, and $z$ components of the orientation vector of a reference MA molecule. 
As defined, the dots on the reference molecule's sphere are colored by position: red for the left and right sides, blue for the top and bottom, and green for the front. 
The orientations of two other investigated MA molecules in the simulation cell (one nearby and one distant from the reference) are also plotted using the color defined by the reference molecule. 
If dots of the same color cluster on the spheres of the other molecules, their orientations are correlated with that of the reference molecule. 
Conversely, if a color is scattered across the sphere, the orientations are uncorrelated.

As shown in the top-left panel of Fig.\ \ref{Orientation_plots_MAPbI3_cell}, in the average cell of the lowest temperature, the orientation of the reference molecule is correlated with that of both nearby and distant MA molecules, indicating strong orientational coupling.
This is consistent with observations by Escorihuela-Sayalero \textit{et al.}\cite{Escorihuela-Sayalero2024, Escorihuela-Sayalero2025} 
This coupling restricts the number of energetically relevant minima.
At higher temperatures and in larger simulation cells, each color is more uniformly distributed for the neighboring molecules, suggesting reduced influence from the reference molecule's orientation. 
In the absence of such coupling, the number of accessible minima is expected to scale exponentially with system size, contributing significantly to the configurational entropy.
As discussed in Sec.\ \ref{SI_orientation_org} of the SI, for other phases and for FAPbI$_3$, the orientation of one molecule can give rise to multiple orientations of surrounding molecules, even at low temperatures and in small simulation cells. 
This strong temperature and cell-size dependence of the orientational entropy in the $\gamma$ phase of MAPbI$_3$ explains the pronounced curvature in the free energy differences shown in Fig.\ \ref{Free_energy_contributions_8fu}, which is even more pronounced in the supercell simulations (see Sec.\ \ref{SI_free_energy_64fu} of the SI).

Remarkably, the temperature dependence of the Helmholtz free energy differences between the phases (shown in the bottom-right panel of Fig.\ \ref{Free_energy_contributions_8fu}) is similar for all three materials, with the exception of the smallest simulation cell of the $\gamma$ phase of MAPbI$_3$, where the configurational free energy is reduced due to hindered orientational freedom of the MA molecules.  
The free energy difference between the $\delta_\text{Cs}$ and $\gamma$ phases increases by approximately 2.4 kJ/mol per 100 K, while for the $\delta_\text{FA}$ and $\gamma$ phases this increase is about 0.9 kJ/mol per 100 K.  
For larger supercells, these values are slightly lower due to additional destabilization of the $\gamma$ phase, as discussed in Sec.\ \ref{SI_free_energy_64fu} of the SI.  
These trends can be used to estimate the transition temperature based on the ground state energy differences.  
However, due to the strong cell dependence, the ground state energy should be evaluated using the average simulation cell at the estimated transition temperature.  
It is also important to note that this apparent similarity in temperature dependence arises from the combined effect of harmonic, anharmonic, and configurational contributions to the free energy.  
These individual components vary significantly between CsPbI$_3$ and the organic MHPs, so caution should be exercised when generalizing this trend.

\section{Conclusions}
MHPs are complex materials that require computationally demanding \textit{ab initio} methods to accurately capture their energies and forces.  
In addition, their structural flexibility gives rise to strong anharmonic effects and a rich landscape of local minima.  
By employing MLPs, we can efficiently reproduce \textit{ab initio}-level energies and forces, enabling MD simulations to compute Gibbs free energies using a TI based methodology.

In this work, we benchmark a range of XC functionals and dispersion corrections for their ability to reproduce the RPA+HF energies and forces of CsPbI$_3$, FAPbI$_3$, and MAPbI$_3$.  
Several XC functionals achieve a reasonable root mean square error (RMSE) in the energies, around 2 kJ/mol pfu.  
However, accurately reproducing the forces proves more challenging.  
The overall RMSE in the forces is approximately 300meV/\AA, but this error is highly element-dependent: the inorganic atoms typically show much smaller force errors (around 100 meV/\AA), while the organic atoms can exhibit errors of up to 1 eV/\AA.  
We attribute these large deviations primarily to internal modes of the organic cations, which are present in all three phases and are thus expected to cancel out when computing free energy differences.  
Based on its balance between accuracy, computational cost, and precision, the MLP was trained on a dataset computed using DFT with the PBE+D3(BJ) functional.

To calculate the Gibbs free energy, we outline a multistep procedure using TI to correct the harmonic free energies. 
The harmonic free energy, previously obtained via molecular dynamics simulations as described in our earlier work,\cite{Braeckevelt2022_CsPbI3} becomes unreliable for materials containing organic molecules due to their rotational degrees of freedom. 
We introduce an intermediate PES an apply REX to mitigate sampling issues at low temperatures, which arise from the presence of multiple local minima. 
The dependence of the Helmholtz free energy on the cell parameters is challenging to correct with high accuracy. 
To reduce this cell dependence, we employ eight different simulation cells for each phase, corresponding to the average cell geometries obtained from NPT simulations at eight different temperatures.

With this TI approach, we find that the $\delta$ phases are enthalpically favored, while the $\gamma$ phase is stabilized by entropy. 
This explains why it is difficult to stabilize the optically active $\gamma$ phase at low temperatures.

The ground state energy favors the $\delta_\text{Cs}$ phase over the $\delta_\text{FA}$ phase, which in turn is more stable than the $\gamma$ phase. 
In contrast, both the harmonic and configurational free energy contributions reverse this stability order, favoring the $\gamma$ phase. 
The predicted transition temperatures for CsPbI$_3$ and FAPbI$_3$ are in good agreement with experimental values, while the transition temperature of MAPbI$_3$ is overestimated. 
Since the free energy differences vary slowly with temperature, even small enthalpic errors can lead to significant deviations in the predicted transition temperatures. 
The free energy difference between the $\delta_\text{Cs}$ and $\gamma$ phases increases by approximately 2.4 kJ/mol per 100 K, whereas the difference between the $\delta_\text{FA}$ and $\gamma$ phases increases by about 0.9 kJ/mol per 100 K. 
For smaller supercells, the harmonic and configurational contributions tend to overestimate the stability of the $\gamma$ phase. 
As a result, simulations with larger supercells exhibit a slightly lower temperature dependence of the free energy difference, shifting the predicted transition temperatures further away from experimental values.

By comparing the different contributions to the free energy for the three materials, we observe that the ground state energy has the largest influence. 
The harmonic free energy also varies between the three materials. 
However, once the configurational free energy is included, the temperature dependence of the total free energy becomes similar across all phases and materials. 
This observation does not hold for the smallest simulation cell of the $\gamma$ phase of MAPbI$_3$. 
In this case, the orientation of a single organic molecule strongly influences the orientation of neighboring molecules. 
In contrast, for larger cells and at higher temperatures, the organic molecules can rotate more freely, resulting in a higher configurational entropy. 
Furthermore, the ground state energy itself is significantly affected by the size and shape of the simulation cell. 
Therefore, in order to reliably estimate the transition temperature based on ground state energy calculations, the energy should be computed using the average simulation cell at the estimated transition temperature. 
Additionally, the organic molecules should exhibit sufficient orientational freedom such that the orientation of one molecule does not strongly constrain the orientations of others.

To the best of our knowledge, this is the first study to compute the Gibbs free energy of both inorganic and organic MHPs with \textit{ab initio} accuracy and minimal approximations. 
Our findings, including the similar temperature dependence of free energy in CsPbI$_3$, FAPbI$_3$, and MAPbI$_3$, the critical role of ground-state energy, and the influence of molecular orientation freedom, provide valuable insights into phase stability across MHP materials. 
Furthermore, the presented methodology enables computational screening for MHPs with enhanced phase stability, offering a valuable tool for guiding materials design and potential industrial applications.

\section*{Conflicts of interest}
\footnotesize
There are no conflicts to declare.

\section*{Data availability}
\footnotesize
The data that support the findings of this study are openly available in Zenodo at https://doi.org/10.5281/zenodo.18108633.

\section*{Acknowledgements}
\footnotesize
The authors acknowledge financial support from iBOF-21-085 PERSIST. 
V.V.S. acknowledges funding from the Research Board of Ghent University (BOF). 
The computational resources and services used in this work were provided by the VSC (Flemish Supercomputer Center), funded by the Research Foundation-Flanders (FWO) and the Flemish Government-department EWI.

\section*{Supporting Information}
\footnotesize
The Supporting Information is provided after the references.

\bibliography{bibliography}

\includepdf[pages=-, fitpaper=true]{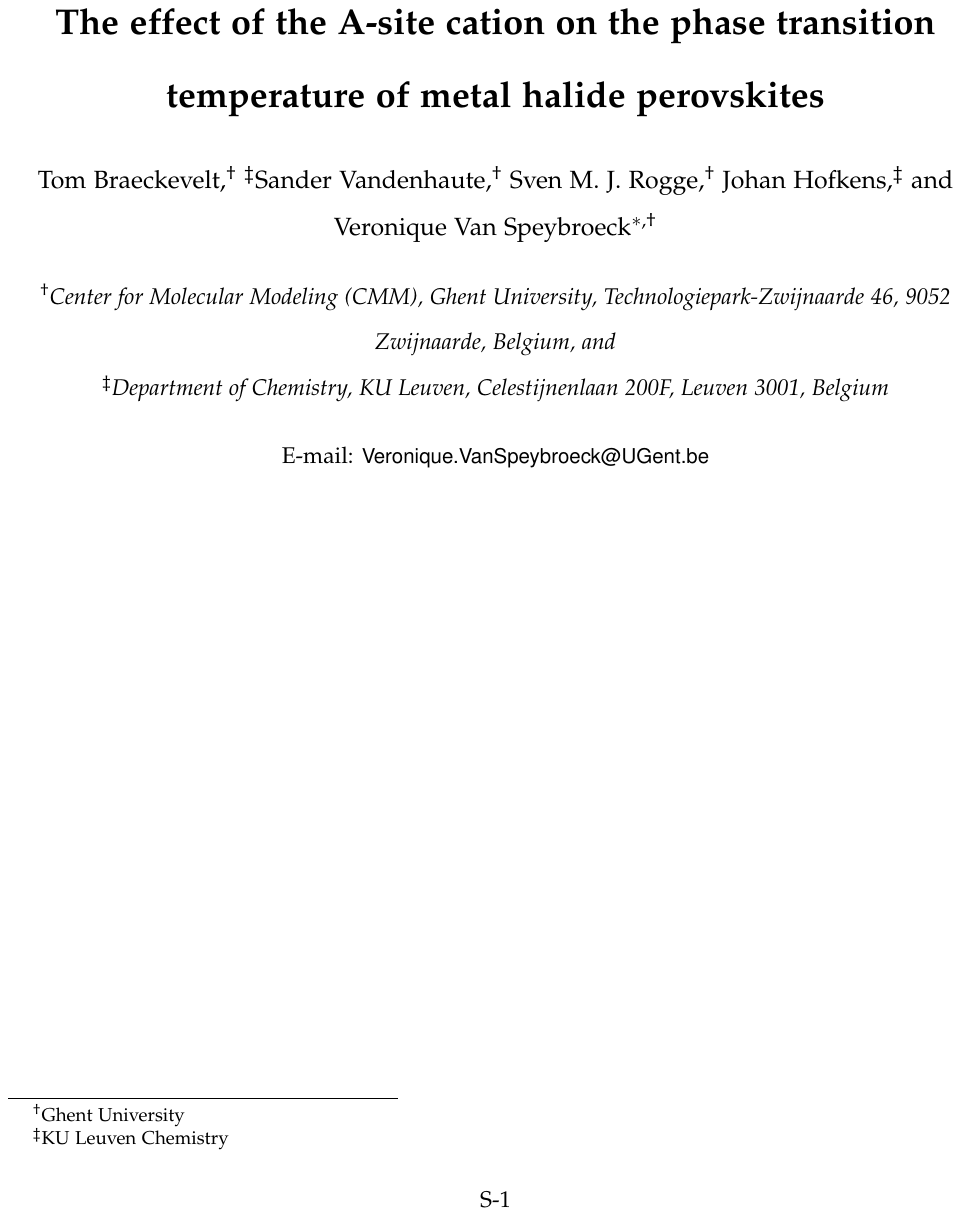}

\end{document}